\documentclass[amsmath,amssymb,pra,aps,showpacs,superscriptaddress,twocolumn]{revtex4-1}
\usepackage{amsmath,amsfonts,amssymb,amsthm,graphics,graphicx,epsfig,bbm}
\usepackage[colorlinks=true,citecolor=blue,linkcolor=blue,urlcolor=blue]{hyperref}
\usepackage[usenames]{color}
\usepackage{graphicx}
\usepackage{subfigure}
\usepackage{amsmath}
\usepackage{epsfig}
\usepackage{dcolumn}
\usepackage{bm}
\usepackage{color}
\usepackage{epstopdf}
\usepackage{amssymb}
\usepackage{amstext}
\usepackage{latexsym}
\usepackage{hyperref}
\usepackage{amsfonts}
\usepackage{psfrag}
\usepackage{xcolor}
\usepackage[normalem]{ulem}
\usepackage{dsfont}
\usepackage{txfonts}

\newcommand{\ket}[1]{\vert #1 \rangle}
\newcommand{\bra}[1]{\langle #1 \vert}
\newcommand{\ketbra}[2]{\vert #1 \rangle \langle #2 \vert}
\newcommand{\braket}[2]{\langle #1 \vert #2 \rangle}

\begin{document}

\title{Creation of quantum error correcting codes in the ultrastrong coupling regime}

\author{T. H. Kyaw}
\affiliation{Centre for Quantum Technologies, National University of Singapore, 3 Science Drive 2, Singapore 117543, Singapore}
\author{D. A. Herrera-Mart\'{i}}
\affiliation{Centre for Quantum Technologies, National University of Singapore, 3 Science Drive 2, Singapore 117543, Singapore}
\affiliation{Racah Institute of Physics, The Hebrew University of Jerusalem, Jerusalem 91904, Israel}
\author{E. Solano}
\affiliation{Department of Physical Chemistry, University of the Basque Country UPV/EHU, Apartado 644, E-48080 Bilbao, Spain}
\affiliation{IKERBASQUE, Basque Foundation for Science, Maria Diaz de Haro 3, 48013 Bilbao, Spain}
\author{G. Romero}
\affiliation{Department of Physical Chemistry, University of the Basque Country UPV/EHU, Apartado 644, E-48080 Bilbao, Spain}
\author{L.-C. Kwek}
\affiliation{Centre for Quantum Technologies, National University of Singapore, 3 Science Drive 2, Singapore 117543, Singapore}
\affiliation{Institute of Advanced Studies, Nanyang Technological University, 60 Nanyang View, Singapore 639673, Singapore}
\affiliation{National Institute of Education, Nanyang Technological University, 1 Nanyang Walk, Singapore 637616, Singapore}

\date{\today}

\begin{abstract}
We propose to construct large quantum graph codes by means of superconducting circuits working at the ultrastrong coupling regime. In this physical scenario, we are able to create a cluster state between any pair of qubits within a fraction of a nanosecond. To exemplify our proposal, creation of the five-qubit and Steane codes is numerically simulated. We also provide optimal operating conditions with which the graph codes can be realized with state-of-the-art superconducting technologies.
\end{abstract}

\maketitle

\section{Introduction} 
Quantum computers promise speedup and robust computational power over their classical counterparts \cite{q_comp,Nielsen2000}. However, their practical realization is still challenging because of susceptibility to errors. Thanks to quantum error correcting codes (QECCs) \cite{Shor1995,Steane1996} and the theory of fault-tolerant quantum computation \cite{Gottesman1998}, these errors can, in principle, be suppressed and corrected in efficient manners. Nowadays, simple quantum error correcting codes that have been experimentally demonstrated are the three-qubit \cite{Chiaverini2004,Reed2012} and four-qubit codes \cite{Lu2008,Bell2014}. Notice that the smallest QECCs capable of correcting both bit-flip and phase errors are the five-qubit and the seven-qubit codes, respectively. With recent advancements in trapped ions \cite{7qubit} and superconducting circuits \cite{JMChow2014} to achieve simultaneous detection of multiple errors, the aim of large scale quantum error correcting codes comes close to reality.

On the road towards realizing quantum graph codes and other complex codes, circuit quantum electrodynamics (cQED) \cite{cQED_architecture1,cQED_architecture2,cQED_architecture3} is a prime candidate for implementing QECCs due to their high level of controllability \cite{Kelly2014,Barends2014,Chen2014} and scalability \cite{Chow2013,Sank2014}. Furthermore, it has been shown both theoretically \cite{Bourassa2009} and experimentally \cite{Niemczyk2010,Pol2010,Baust2014} that a flux qubit galvanically coupled to a coplanar waveguide resonator reaches the ultrastrong coupling (USC) regime \cite{Ciuti2005} of light-matter interaction, where the qubit-resonator coupling strength $g$ is comparable to the resonator frequency $\omega$; i.e., $0.1\lesssim \! g/ \omega\lesssim \! 1$. This coupling regime enables direct application of ultrafast two-qubit gates \cite{USC_gate} between a pair of qubits inside the resonator. We aim to realize the USC gate in between any pair of qubits within a resonator, to realize complex quantum codes for quantum error correction schemes in a scalable manner.

Here, we show how to construct two QECCs, the five-qubit code \cite{Nielsen2000} and the Steane code \cite{Steane1996}, in a cQED architecture operating at the USC regime. We construct them by sequentially performing ultrafast controlled phase gates [$U_{\mathcal{CZ}}=\textrm{diag}(1,1,1,-1)$] between any two physical qubits to encode one logical qubit. Ultrafast gate time and high fidelity response of the superconducting circuit might ensure very low errors incurred at the logical qubit level (see \textit{``Errors and decoherence model"} section). We believe our scheme could be used to mediate interactions between logical qubits and perform protected quantum computations in a measurement-based manner \cite{1way}. In addition, our proposal may pave a way to construct various types of QECC applications \cite{Ozeri2013, Arrad2014}. 

\section{Superconducting circuit design} 
\begin{figure}[t]
\centering
\includegraphics[scale=0.5]{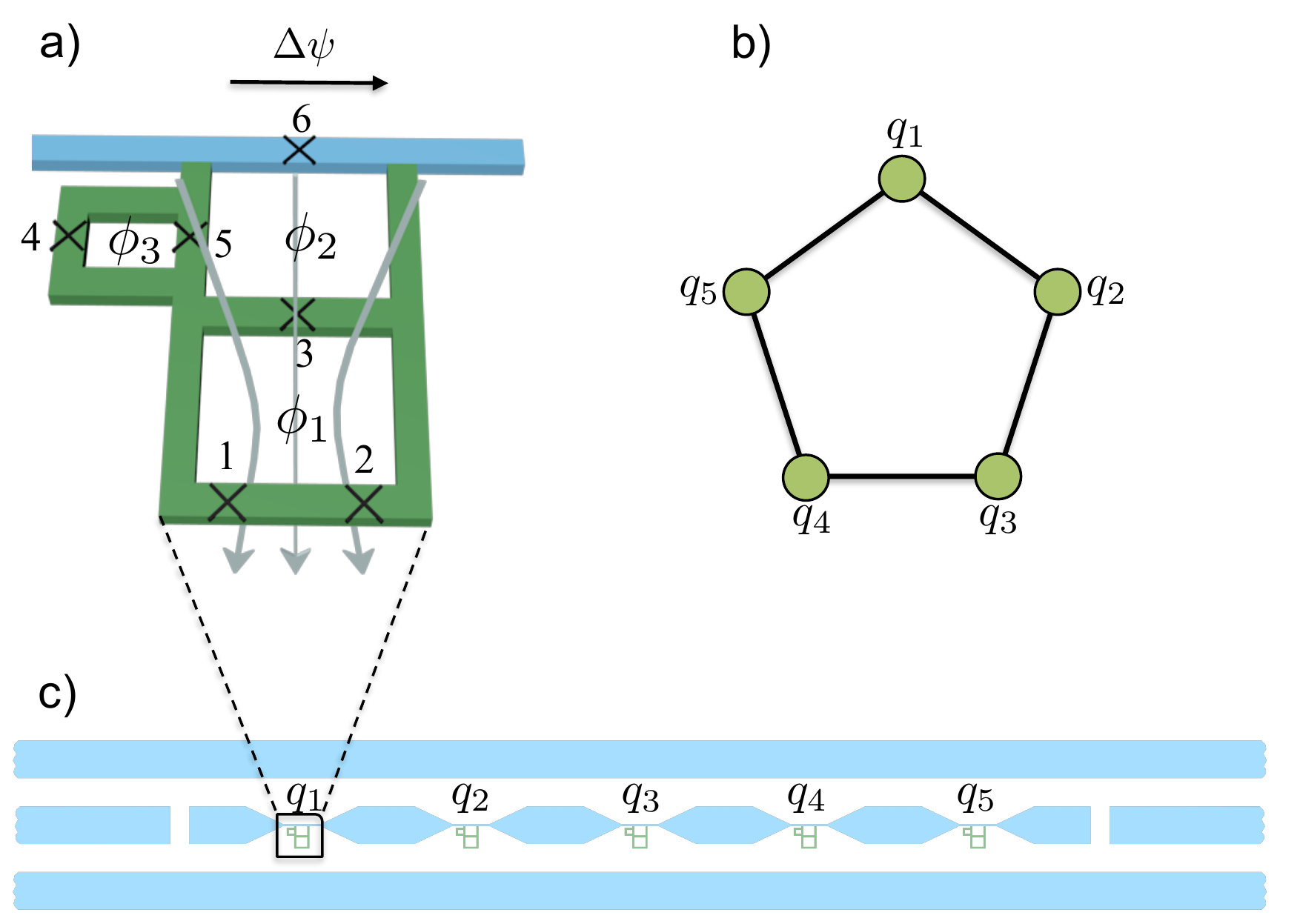}
\caption{(color online). (a) Schematic of a flux qubit, denoted by the Josephson junctions 1, 2 and 3. By varying the frustration parameter $f_3$, attained by an applied magnetic flux passing through the loop composed of the JJ$_4$ and JJ$_5$, the coupling between the qubit and the resonator can be tuned at will. This is a crucial aspect of our superconducting qubits design in order to realize cluster states in an ultrafast timescale.
(b) A five-qubit cluster state. Each black bond represents the pair-wise cluster state generation mechanism $U_{\mathcal{CZ}}^{ij}$ between $i$th and $j$th physical qubits (green circles) that are initially prepared in the $\ket{+}=(\ket{g}+\ket{e})/\sqrt{2}$ state.
(c) An array of five USC qubits embedded in a resonator to obtain the five-qubit quantum error correcting code. 
}
\label{Fig1}
\end{figure}

In order to achieve ultrafast quantum gate operations in between any two physical qubits, we consider a superconducting flux qubit \cite{flux_qubit} (see Fig.~\ref{Fig1}), which consists of six Josephson junctions (JJs). Each JJ is denoted with a cross, and it is galvanically coupled \cite{Niemczyk2010,Pol2010} to a coplanar waveguide resonator (CWR) by means of the coupling junction, JJ$_6$. This qubit configuration provides a tunable qubit-resonator coupling strength \cite{USC_gate,Peropadre2010}, where the flux qubit potential energy is defined by junctions 1, 2 and 3
\begin{eqnarray}\label{Energy_potential}
\frac{U_q}{E_{J}}=&-&\lbrace \cos\varphi_1 + \cos\varphi_2 + \alpha \cos(\varphi_2 - \varphi_1 + 2\pi f_1)\\
	&+& 2\beta \cos (\pi f_3)\cos[ \varphi_2 -\varphi_1 +2\pi (f_1 -f_2 +f_3 /2)+\varphi_x]\rbrace.\nonumber
\end{eqnarray} 
In the configuration of Fig.~\ref{Fig1}(a), each JJ contributes an energy $\mathcal{E} (\varphi_i)=-E_{J_i} \cos(\varphi_i)$, where $E_{J_i}$ and $\varphi_i$ are the Josephson energy and the gauge-invariant superconducting phase difference across the $i$th junction. We assume $E_{J_1}=E_{J_2}=E_{J}$, $E_{J_3}=\alpha E_{J}$, $E_{J_4}=E_{J_5}=\beta E_{J}$, and the quantization rule for each closed loop, i.e., $\sum_{j} \varphi_j =2\pi f_i +2\pi n$ where $f_i = \phi_i /\Phi_0 $ is the frustration parameter, $\Phi_0 =h/2e$ is the flux quantum. In addition, $\varphi_x$ is the phase slip shared by the resonator and the $f_2$ loop (see appendix), and $n$ is an integer multiple. Near the symmetry point, i.e., $\phi_1\approx\Phi_0/2$, the flux qubit potential (\ref{Energy_potential}) can be effectively truncated to a two-level system with frequency $\omega_q=\sqrt{\Delta^2+\varepsilon^2}/\hbar$. Here, $\Delta$ is the qubit energy gap and $\varepsilon = 2I_p(\phi_1-\Phi_0/2)$ with $I_p$ being the persistent current.

We propose to construct two QECCs, the five-qubit code and the Steane code, by considering the cQED architecture shown in Fig.~\ref{Fig1}(c), where the tunable flux qubits are uniformly distributed along a CWR \cite{supplemental}. In this configuration, we assume that each flux qubit acts as a small perturbation to the resonator due to the presence of JJ$_6$'s. This condition is satisfied when the inductance of the JJ$_6$'s is much smaller than the sum of inductances belonging to the loop threaded by the external flux $\phi_2$ (see Fig.~\ref{Fig1}(a)). In this case, most of the current will flow along the microwave resonator~\cite{Bourassa2009}, a condition that has already been achieved in experiments for implementing the USC regime~\cite{Niemczyk2010,Baust2014}. In addition, each JJ$_6$ will introduce extra boundary conditions on the resonator that together with open boundary conditions at the resonator edges, will allow us to define an eigenmode structure (see appendix). A case of interest occurs when JJ$_6$'s operate in the linear response regime where the Josephson energy $E_J$ is much larger than the capacitive energy $E_C$. This leads to a non-linear resonator spectrum where each harmonic presents a manifold $\mathcal{M}$, whose number of bosonic modes corresponds to the number of qubits embedded across the resonator \cite{Leib2014}. Under the aforementioned conditions, we are led to an effective system Hamiltonian (see appendices for detailed derivations) which reads
\begin{eqnarray}\label{effH}
\mathcal{H} &=& \frac{\hbar}{2}\sum^{N}_{j=1}\omega^j_q\sigma^j_z+\hbar\sum_{\ell\in\mathcal{M}}\omega_{\ell}a^{\dag}_{\ell}a_{\ell}+\nonumber\\
&&\hbar\sum^{N}_{j=1}\sum_{\ell\in\mathcal{M}}g_j(c^j_x\sigma^j_x+c^j_z\sigma^j_z)(a_{\ell}+a^{\dag}_{\ell}),
\end{eqnarray}
where $\omega_q^j$ is the $j$th qubit frequency, $\sigma_{z,x}^j$ are the Pauli matrices, $\omega_l$ is the frequency of the $\ell$th resonator mode belonging to the manifold $\mathcal{M}$, $a^\dagger _\ell(a_\ell)$ is the creation (annihilation) operator of $\ell$th resonator mode, and the coefficients $c_{x}^j$ and $c_{z}^j$ are functions of the system parameters $\alpha, \beta, f_1$, and $f_2$ \cite{USC_gate}, satisfying the condition $|c^{j}_x|^2+|c^{j}_z|^2=1$ for $\forall j$. The coupling strength $g_j\propto 2E_J \beta \cos(\pi f_3)/\hbar$, depends on the external magnetic flux $\phi_3$, and $N$ is the total number of qubits present in the resonator. We note that different coupling strengths appear due to the spatial dependence of the distribution of flux qubits along the resonator \cite{supplemental}. Moreover, it has been shown that different frequencies belonging to a specific manifold $\mathcal{M}$ become degenerate  ($\omega_{\ell}=\omega$) \cite{Leib2014} for a specific value of the plasma frequency $\omega_p=1/\sqrt{C_JL_J}$ associated with the coupling junctions JJ$_6$, where $C_J$ is the Josephson capacitance, $L_J=\varphi^2_0/E_J$ is the Josephson inductance, and $\varphi_0=\Phi_0/2\pi$ is the reduced flux quantum.   

It is noteworthy that coefficients $c_x ^j$ and $c_z ^j$ can be manipulated by means of the external flux $\phi_1 ^j$ as shown in Fig.~\ref{Fig2}(a,b) for a given junction size $\alpha=E_{J_3}/E_J$. Here, it might be possible to tune the transversal coupling where $c_x\to1$ and $c_z\to0$, or the longitudinal coupling where $c_x\to0$ and $c_z\to1$. The latter becomes an essential condition for generating pairwise cluster states. The numerical simulation of coefficients $c_x$ and $c_z$ in Fig.~\ref{Fig2}(a,b) has been performed by diagonalizing the flux qubit potential, Eq.~(\ref{Energy_potential}), and truncating it to the two lowest energy levels \cite{USC_gate}. This allows us to evaluate numerically matrix elements of the persistent-current operator $I_{jk}\propto\langle j|\sin(\varphi_2 ^j -\varphi_1 ^j+2\pi (f_1 ^j-f_2 ^j +f_3 ^j/2))|k\rangle$ in the basis of the effective two-level system to obtain $\sin(\varphi_2 ^j -\varphi_1 ^j+2\pi (f_1 ^j-f_2 ^j +f_3 ^j/2))=\sum_{\nu=0,x,y,z}c_\nu\sigma_\nu$, with $\sigma_0=\mathbbm{1}$ being the identity operator (see appendices).

\section{Pairwise cluster state generation} 
\begin{figure}[t]
\centering
\includegraphics[width=0.45\textwidth]{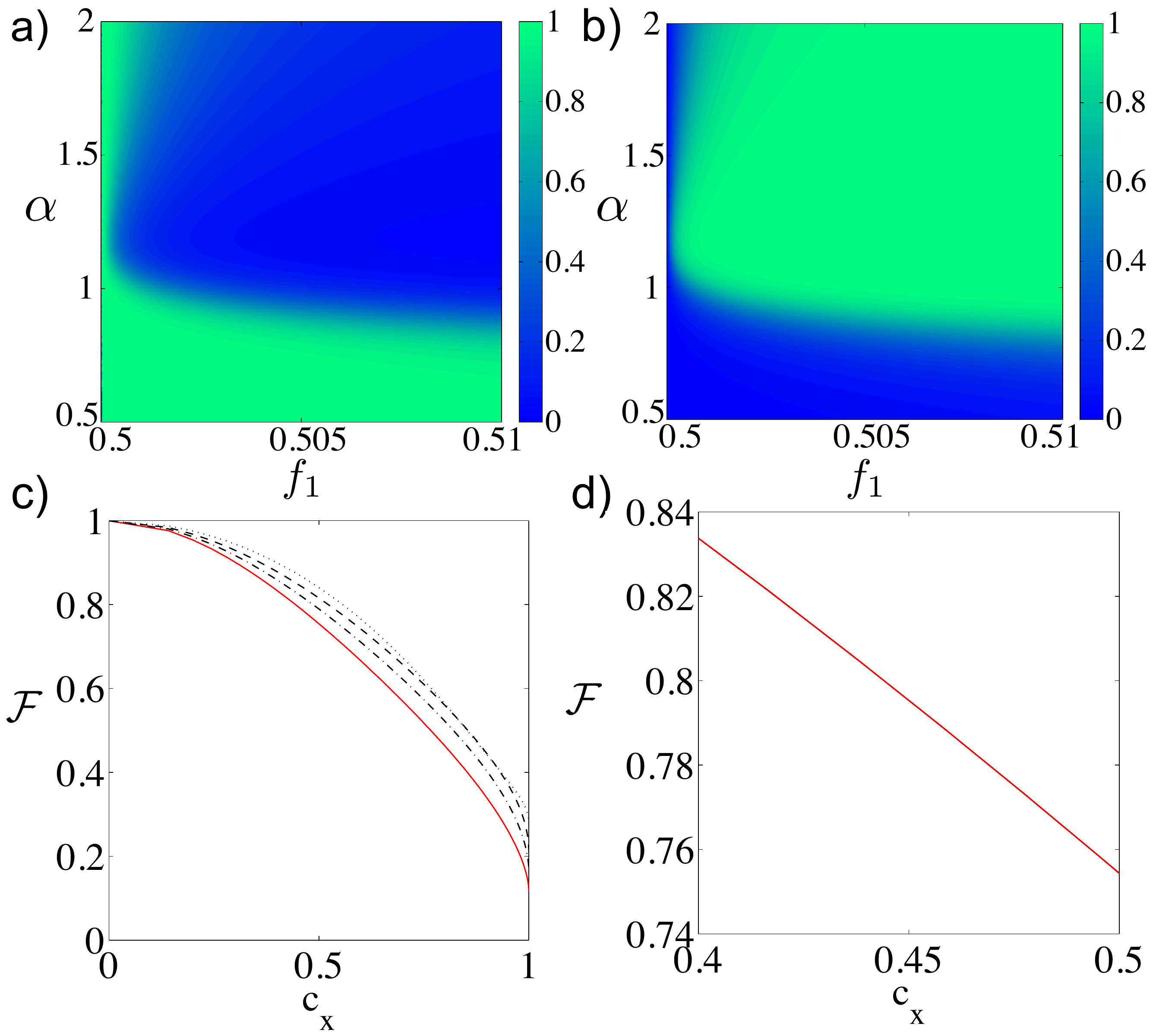}
\caption{(color online). Coupling coefficients (a) $c_x$ and (b) $c_z$ in Eq.~(\ref{effH}) as a function $\alpha$, that is the size of junction JJ$_3$, and the frustration parameter $f_1=\phi_1/\Phi_0$. In this simulation we have considered $E_J/h=221$~GHz and the capacitive energy of junctions as $E_C=E_J/32$. (c) Fidelity of achieving the desired controlled-phase gate $U_{\mathcal{CZ}}$, in the presence of non-zero transversal coupling strength $c_x$, while the red solid line, the black solid line, the dot-dashed line, the dashed line and the dotted line represent the case when the resonator field is a thermal state at 15 mK, a vacuum state, a coherent state with amplitude $\gamma=0.25$, a coherent state with $\gamma=0.5$ and a coherent state with $\gamma=1$, respectively. (d) The enlarged figure of (c) shows the red and black solid lines overlap each other, indicating that the vacuum and the thermal states behave the same when considering two bosonic modes.
}
\label{Fig2}
\end{figure}
\begin{figure}[b]
\begin{center}
\includegraphics[scale=0.23]{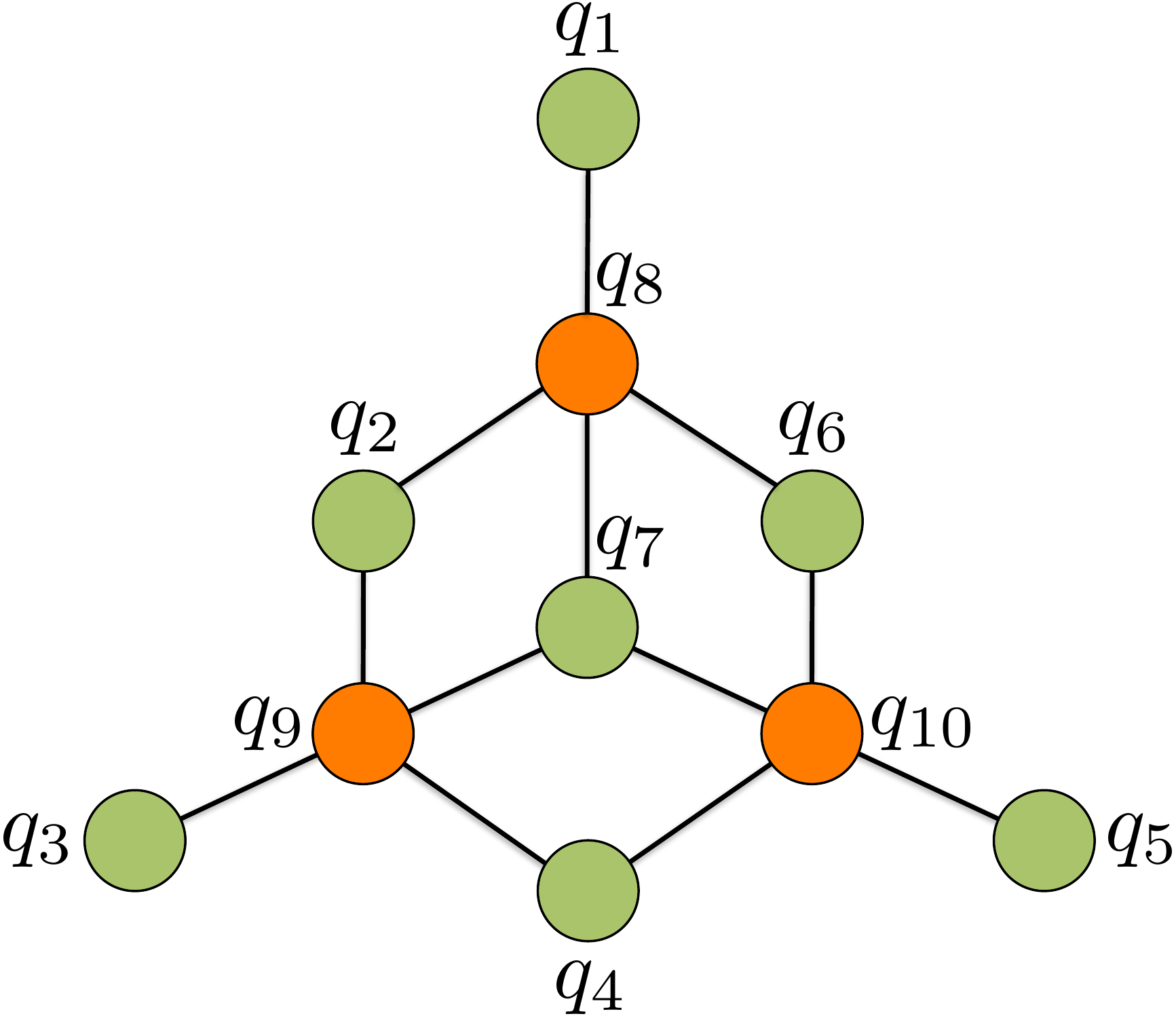}
\caption{(color online). A ten-qubit cluster state emerging from the Steane code after appropriate projective measurements on the physical qubits $q_8, q_9$ and $q_{10}$ (orange circles). Each black bond represents the pair-wise cluster state generation mechanism $U_{\mathcal{CZ}}^{ij}$ between the $i$th and $j$th physical qubits (green circles) that are initially prepared in the $\ket{+}=(\ket{g}+\ket{e})/\sqrt{2}$ state. 
}
\label{Fig3}
\end{center}
\end{figure}

 
A cluster state between any $i$th and $j$th qubits can be readily generated in four steps. Firstly, the system is cooled down to reach its ground state in the USC regime. Secondly, both qubits are adiabatically addressed with external fluxes that vary linearly in time $\phi^k_3=\bar{\phi}_0+\Delta\phi t/T$ where $k\in \{i,j\}$, with $\bar{\phi}_0$ an offset flux, $\Delta\phi$ a small flux amplitude, and $T$ the total evolution time. In this case, the coupling strength of each qubit reaches the strong coupling regime described by the Jaynes-Cummings model \cite{JCmodel} such that the system is prepared in the state $\ket{\psi_G}=\ket{g}^{\otimes N}\otimes\ket{0}^{\otimes N}$, where $\ket{g}$ and $\ket{0}$ stand for the ground state of the qubit and the vacuum state for each mode in $\mathcal{M}$, respectively. The validity of this initialization process can be proven numerically (see Section~\ref{Sec:4}). Thirdly, each qubit is then addressed with a classical microwave signal, sent through the cavity, to be prepared in the state $\ket{+}=(\ket{g}+\ket{e})/\sqrt{2}$ which is an eigenstate of $\sigma_x$, while all the remaining $N-2$ qubits are far off-resonant with respect to the driving frequency. At this stage all qubits should dispersively interact with the modes within the manifold $\mathcal{M}$ such that there is no exchange of excitations. This task might be carried out at a degenerate regime of the bosonic manifold, $\omega_{\ell}=\omega$ \cite{Leib2014}. At last, the external magnetic fluxes $\phi^k_3$ are swiftly tuned to reach the USC coupling strength within a subnanosecond timescale. During these four steps, the magnetic fluxes $\phi^{k}_1$ should be tuned to reach the longitudinal qubit-resonator coupling. After interacting with the collective resonator modes, the system evolution operator takes the form \cite{supplemental, Wang2010}
\begin{align}
U(t)=U_0 (t) e^{i\xi^2 M(\omega t - \sin(\omega t))}\prod_{\ell} e^{-i\omega t a_{\ell} ^\dagger a_{\ell}} \mathcal{D}_{\ell}\big[\kappa(t)\big],
\label{evolU}
\end{align}
where $\mathcal{D}_{\ell}[\kappa(t)]=\exp [\kappa(t) a_{\ell}^{\dag}-\kappa^{*}(t) a_{\ell}]$ is the displacement operator associated with the $\ell$th bosonic mode within the manifold $\mathcal{M}$. In addition, $\xi=\sum_{j=1}^{N}\kappa_j \sigma_z^j$ with $\kappa_j=g_j/\omega$, $M$ stands for the number of degenerate bosonic modes $b_{\ell}$, and the unitary $U_0(t)=\exp(-it \sum_{j=1}^{N} \frac{\omega_q ^j}{2}\sigma_z ^j)$. After the evolution time $T=2\pi n/\omega$, we have performed the desired controlled phase gate operation between the qubits
\begin{eqnarray}
U_{\mathcal{CZ}}&=& \mathcal{U}\times \exp\left[-\frac{i\pi}{4}(\sigma_z ^i +\sigma_z ^j) \right] \\
&\times& \exp\left[4\pi i M\left((\kappa_i ^2 +\kappa_j ^2)\frac{\mathbbm{1}}{2}+\kappa_i \kappa_j \sigma_z ^i \sigma_z ^j \right) \right],\nonumber
\end{eqnarray}
where $\mathcal{U}=\exp \left[\frac{-i\pi}{4} \left[\Big(\frac{4\omega_q ^i -\omega}{\omega}\Big)\sigma_z ^i +\Big(\frac{4\omega_q ^j -\omega}{\omega}\Big)\sigma_z ^j \right]\right]$. The resultant state incurs an extra global phase due to the presence of $\mathcal{U}$, which is unavoidable since it is formidable by construction to tune a desired qubit frequency, via the external flux $\phi_1$, without affecting the longitudinal and transversal coupling strengths (see Fig.~\ref{Fig2}(a) and (b)). To achive maximum gate fidelity, we require both $\kappa_i ^2 +\kappa_j ^2 =\frac{1}{8nM}$ and $\kappa_i \kappa_j =\frac{1}{16nM}$. That means the two coupling strengths need to satisfy $\kappa_i + \kappa_j =\frac{1}{2\sqrt{nM}}$. The operational gate time is estimated to be $T=2\pi/\omega \sim 0.2$ns if the collective mode frequency is $\omega=2\pi \times 5~\text{GHz}$, which implies a ratio $g_j/\omega=1/(4\sqrt{2})\approx 0.17$ belonging to the USC regime. The latter has already been demonstrated in a recent experiment~\cite{Baust2014}. As soon as the two qubits are entangled, they are immediately detuned from the resonant frequency so that we may repeat the same procedure for other qubit pairs to arrive at a specific quantum error correcting code, that being the five-qubit code (see Fig. \ref{Fig1}(b)) or the Steane code (see Fig. \ref{Fig3}).

\subsection{Five-qubit code} 
To demonstrate our ultrafast cluster state generation scheme, we create the five-qubit code which is the smallest QECC that protects against an arbitrary error on a single qubit encoded state \cite{Nielsen2000}. We recall that a cluster state is a common eigenstate of stabilizer operators $K_i = X_i\bigotimes_{j\in nb(i)} Z_j$, where $X_i =\sigma_x^i$, $Z_i=\sigma_z^i$ and $nb(i)$ means neighbours of the $i$th qubit. Since the stabilizer operators form a group $\ket{\psi} = K_i\ket{\psi} = K_iK_j\ket{\psi}$, it is possible to define $S'_i = K_iK_{i+1\hspace{0.05cm}\textrm{mod}\hspace{0.05cm}5}$ and logical operators $\bar{X} = K_5$ and $\bar{Z}=Z_1Z_2Z_3Z_4Z_5$, from which it follows that the five-qubit cluster state is equivalent to the five-qubit code via local unitary transformation $U= \bigotimes_i S_i H_i$, where $S_i$ $(H_i)$ is the phase (Hadamard) gate (see Ref.~\cite{David2013}). Therefore, we create the five-qubit cluster state shown in Fig.~\ref{Fig1}(b) by applying the pairwise cluster state generation mechanism $U_{\mathcal{CZ}}^{ij}$. The resultant state is
\begin{align}
\ket{\Psi_5} =  U_{\mathcal{CZ}}^{15} U_{\mathcal{CZ}}^{54} U_{\mathcal{CZ}}^{43} U_{\mathcal{CZ}}^{32} U_{\mathcal{CZ}}^{21} \ket{+}^{\otimes 5},\label{5qubitcode}
\end{align}
after an evolution time $\tau_5 =10\pi/\omega$. After local operations acting on individual qubits, we achieve the five-qubit code.

\subsection{Steane Code} 
The Steane code \cite{Steane1996} can also be constructed in a manner similar to that of the five-qubit code, but from a cluster state of ten qubits as shown in Fig.~\ref{Fig3}. We require seven stabilizer operators, among ten possible operators, in the combination of operators that commute with $X_8$, $X_9$ and $X_{10}$. It can easily be checked that measuring the orange colored qubits in the $X$ basis leaves the remaining seven qubits in the desired code state \cite{Nielsen2000}. 
With twelve $U_{\mathcal{CZ}}^{ij}$ gates followed by three parallel measurements within an evolution time of $\tau_7 = 24\pi/\omega$, we achieve the Steane code
\begin{equation}
\ket{\Psi_7} =  \bra{+}_{10}\bra{+}_9\bra{+}_8\prod_{k\in E} U_{\mathcal{CZ}}^k\ket{+}^{\otimes 10},\label{steanecode}
\end{equation}
with $E$ representing the set of all the black colored bonds in Fig. \ref{Fig3}.

\begin{figure}[t]
\centering
\includegraphics[scale=0.35]{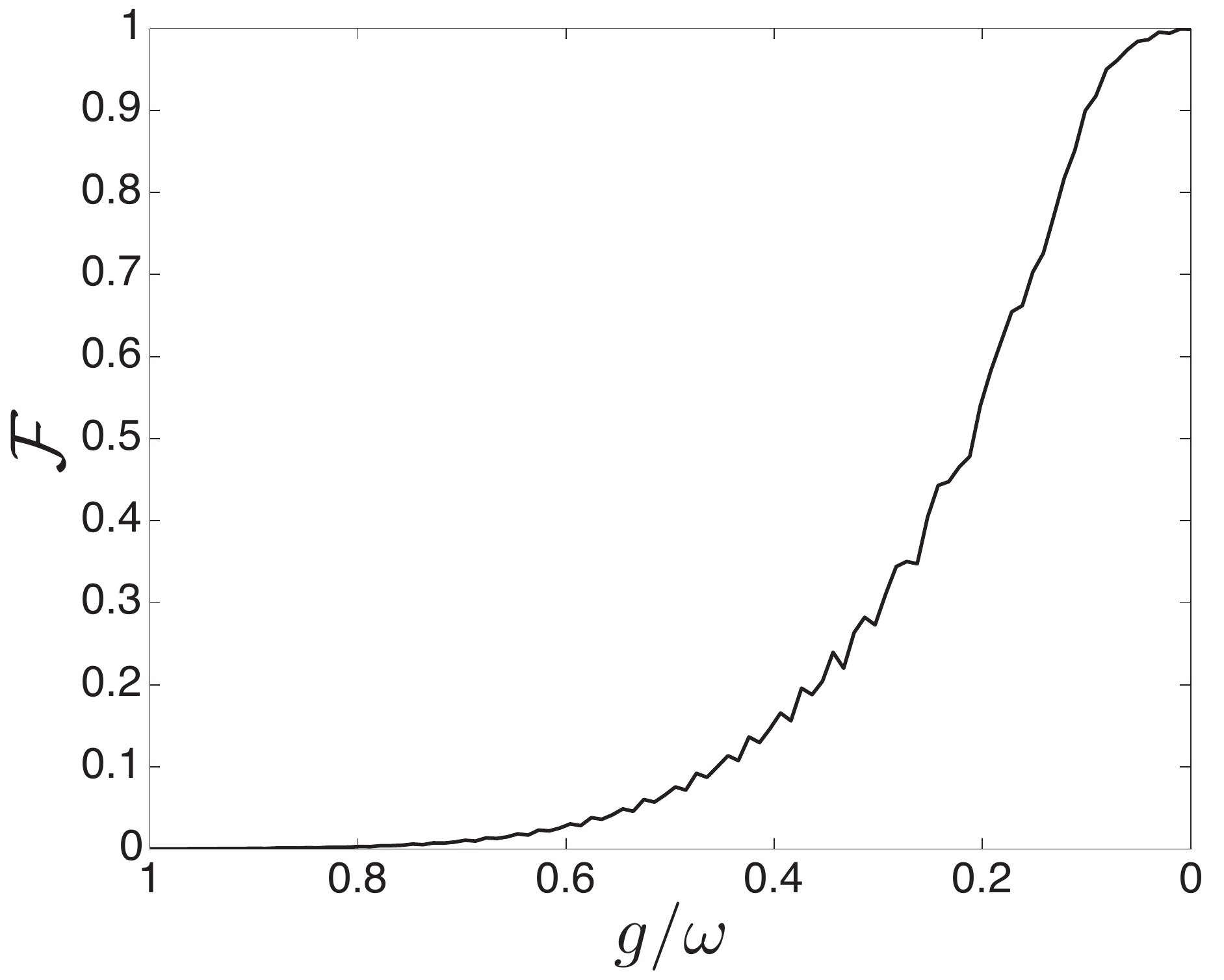}
\caption{Fidelity between the desired Jaynes-Cummings ground state $\ket{\psi_{\text{JC}}}=\ket{gg}\otimes \ket{00}$ and the instantaneous state $\ket{\psi(t)}$ during an adiabatic switch-off process, provided that an initial state of the evolution is $\ket{\psi_G}$, ground state of the quantum Rabi model. In another words, $\mathcal{F}=|\langle \psi_{\text{JC}}|\psi(t)\rangle|^2$ is plotted against $g/\omega$.}
\label{USC-JC}
\end{figure}

\section{Validity of the rotating wave approximation during state initialization}
\label{Sec:4}

To initialize our system for the ultrafast cluster state creation, we intend to cool down the entire system to its ground state. By design, our qubits are galvanically coupled to the resonator. Hence, we expect them to reach the ultrastrong coupling (USC) regime at the end of the cooling process. Afterwards, the coupling strength $g$ is adiabatically switched off till it reaches the Jaynes-Cummings (JC) regime. As a consequence, the ground state of the quantum Rabi model, $\ket{\psi_G}$, which has just been prepared by cooling, is adiabatically mapped to the JC ground state, $\ket{\psi_{\text{JC}}}=\ket{g}^{\otimes N}\otimes \ket{0}^{\otimes N}$, where $\ket{g}$ is the ground state of a qubit, $\ket{0}$ is the bosonic mode vacuum, and $N$ is the number of qubits and bosonic modes present in the resonator. At this moment, our system is ready for the ultrafast cluster state creation process.

To illustrate our protocol, we consider two qubits embedded in a resonator with two modes and simulate the aforementioned adiabatic process. Fig. \ref{USC-JC} shows the fidelity plot of the JC ground state $\ket{\psi_{\text{JC}}}=\ket{gg}\otimes \ket{00}$ and the instantaneous state $\ket{\psi(t)}$ during an adiabatic switch-off process, given the initial state $\ket{\psi_G}$, that is, the ground state of the quantum Rabi model. The initialization process via adiabatic switch-off takes $T=250/\omega= 50$ ns, if we take the resonator frequency to be at $\omega=5$ GHz. Unit fidelity at the end of the adiabatic evolution ascertains that the the rotating wave approximation is consistent in this context, while the extension to a large number of qubits and bosonic modes is straightforward.
\begin{figure}[b]
\begin{center}
\includegraphics[scale=0.4]{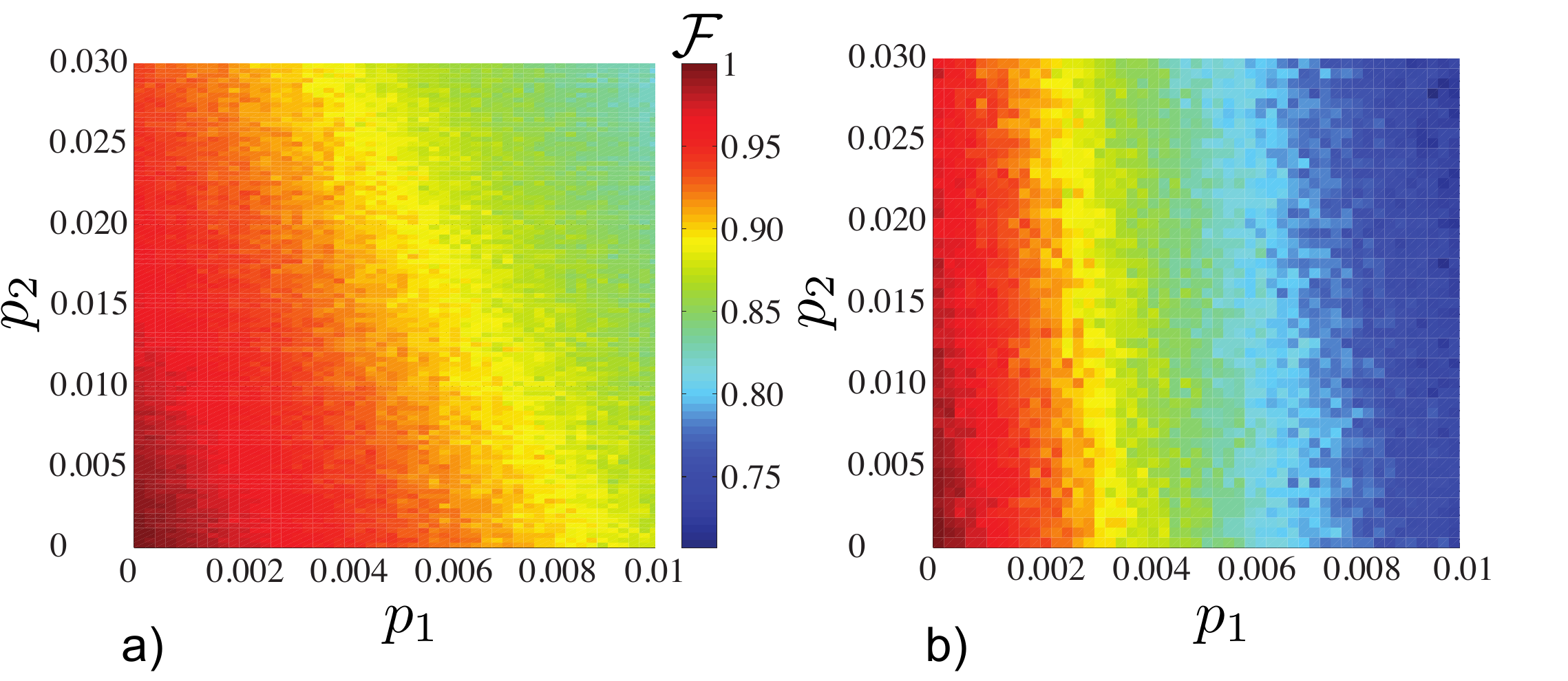}
\caption{(color online). Monte Carlo simulation results. Fidelity of (a) the five-qubit code, where the average fidelity value is taken over 5000 runs and (b) the Steane code, where the average fidelity value is taken over 1000 runs is plotted against single-qubit gate error probability $p_1$ and two-qubit gate error probability $p_2$.
}
\label{Fig5}
\end{center}
\end{figure}

\section{Errors and decoherence model} 
The pairwise cluster state generation mechanism assumes that coupling coefficients $c_x^j = 0$ in the effective Hamiltonian, Eq.~(\ref{effH}). That means we specifically require the longitudinal couplings. However, there might be some residual non-zero transversal couplings in a physical implementation. Whenever this is the case, i.e., $c_x^j \neq 0$, the performance of the ultrafast gate $U_{\mathcal{CZ}}$ is affected, depending on the amount of residuals. In order to see the gate performance with presence of the transversal couplings, we have performed numerical simulations for the dynamics governed by Eq. (\ref{effH}) for the simplest scenario of two-qubits and two bosonic modes belonging to the manifold $\mathcal{M}$. In Fig.~\ref{Fig2}(c,d), we show the optimal operating conditions to obtain maximum gate fidelity. In particular, we plot the fidelity $\mathcal{F}=|\braket{\psi_F}{\psi}|^2$ between the expected final two-qubit state $\ket{{\psi}_F} =\frac{1}{\sqrt{2}}(\ket{e,+}-\ket{g,-})$, with $\ket{\pm}=(\ket{g}\pm\ket{e})/\sqrt{2}$, and the state $\ket{\psi}$ after the pairwise gate operation has been performed with an initial state $\ket{\psi_0}=\ket{+,+}\bra{+,+}\otimes \rho_{F}$ along various values of $c_x$. Notice that in calculating the fidelity $\mathcal{F}$ we have traced out the cavity degrees of freedom. Here, $\rho_{F}$ is the cavity field being a thermal state at 15 mK (red solid line), a vacuum state (black solid line), a coherent state with amplitude $\gamma=0.25$ (dot-dashed line), a coherent state with $\gamma = 0.5$ (dashed line) and a coherent state with $\gamma = 1$ (dotted line), respectively. Even though the presence of vacuum, thermal or coherent state inside the resonator at near $c_x \ll 1$ does not affect much of the gate performance, we note that a coherent state field in the resonator has clear advantage over the true vacuum field. In particular, we observe improvement in the gate fidelity when the resonator field becomes closer to the classical field, i.e., for a coherent state amplitude $\gamma \rightarrow 1$. 

In addition to imperfection of the cavity initial state and coupling strengths, we expect our system to be exposed to thermal noise from the control lines and crosstalks between physical qubits. We model these effects as uncorrelated depolarizing noise which follows otherwise perfect gates, and estimate the fidelity of the final states by performing Monte Carlo simulations for generation of the two QECCs. Also, we consider measurement error of $p_m =0.01$ \cite{Sank2014} for the case of the Steane code. At the end, the collective state of the logical qubit associated with the graph code can be written as $\rho = \mathcal{F}\ketbra{\Psi_\nu}{\Psi_\nu}+ (1-\mathcal{F})\mathbb{I}/2^\nu$, where $\mathcal{F}$ is the fidelity of attaining the five-qubit code ($\nu=5$ and see Fig. \ref{Fig5}(a)) or the Steane code ($\nu=7$ and see Fig. \ref{Fig5}(b)).

\section{Summary and Discussions} 
To summarize, we have proposed a possible realization of the five-qubit and the Steane codes in an array of superconducting circuits galvanically coupled to a coplanar waveguide resonator that mediates two-qubit interactions. The system operates in the USC regime, in which two-qubit gates of subnanosecond timescale are demonstrated. At this timescale, it is strenuous for the gate errors to be limited by the coherence time of the qubit and the resonator in the galvanic configuration \cite{Niemczyk2010,Pol2010}, whose rough estimation is $10-100$ ns and $160-500$ ns, respectively \cite{private}. However, recent randomized benchmarking techniques in circuit QED technologies \cite{Chow2013,Barends2014} have shown that the error per gate can be reduced to about $0.5\% $. This precedent might encourage the realization of our approach, in which fidelities in excess of $75\%$ could be achieved. Also, imperfect measurements are significant sources of errors in the construction of cluster states. However, extremely fast measurements with $99\%$ fidelity have been demonstrated in Ref. \cite{Sank2014}. Moreover, in the light of current developments of large microwave cavity arrays, and following ideas from freely scalable quantum technologies developed in Ref. \cite{Nickerson2014}, one may think of scaling up our system to a two-dimensional array with nearest-neighbor coupling between cavities mediated by superconducting quantum interference devices. As already established in \cite{Nickerson2014}, the scaling up to large architectures does not imply increasing the number of physical qubits inside a unique device; instead, it has been proven that linking cells to one another via noisy channels is fault tolerant if entanglement purification is performed with high fidelity. Thus, we believe our proposal, with all the advanced technologies in the superconducting circuits, might pave a promising avenue for implementing large-scale QECCs or topological codes \cite{TEC1,TEC2,TEC3,TEC4,TEC5} in ultrafast timescale. 

\begin{acknowledgments}
This work was supported by the National Research Foundation, the Ministry of Education, Singapore; Spanish MINECO FIS2012-36673-C03-02; UPV/EHU UFI 11/55; Basque Government IT472-10; SOLID, CCQED, PROMISCE, and SCALEQIT European projects.
\end{acknowledgments}

\appendix

\section{A superconducting coplanar waveguide resonator interrupted by a series of uniformly spaced flux qubits}
\begin{figure*}
\centering
\includegraphics[scale=0.45]{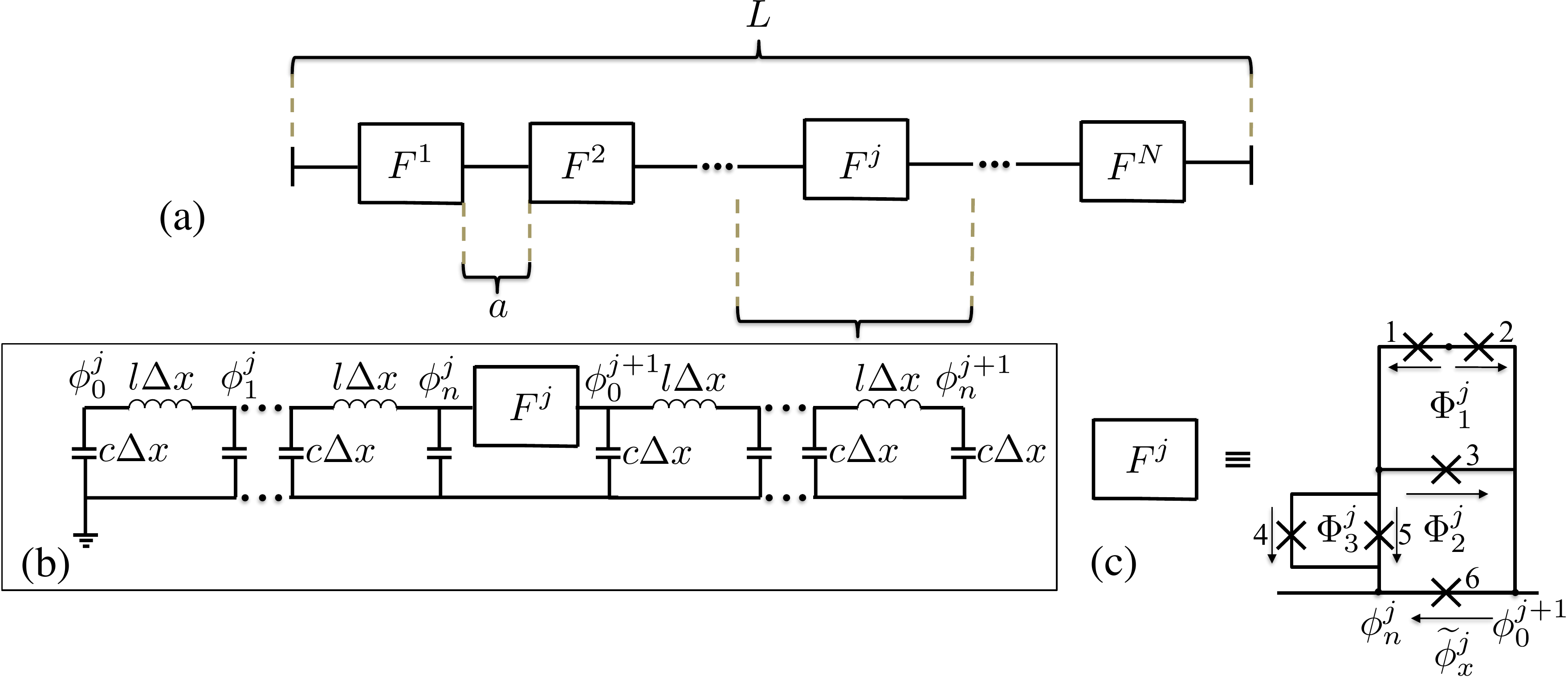}
\caption{(a) A coplanar waveguide resonator (CWR) of length $L$ interrupted by $N$ identical and uniformly distributed flux qubits. 
(b) Lumped-element circuit model for a portion of the CWR, encompassing flux qubit $F^j$.
(c) Schematic of the flux qubit. Numbers 1-6 with a cross sign label Josephson junctions, arrows refer to voltage drop between any two nodes, and $\Phi_{1,2,3}$ are external magnetic fluxes passing through each loop. Here, $\phi$ means a node variable and $\widetilde{\phi}$ means a branch variable, respectively.
}
\label{CPWR}
\end{figure*}
To arrive at the effective Hamiltonian, Eq. (\ref{effH}) of the main text, we consider $N$ identical flux qubits labeled as $F$'s that are uniformly distributed across a CWR (see Fig. \ref{CPWR}(a,b)). Two ends of the resonator are open-circuited, while the CWR supports one-dimensional current-charge waves with phase-velocity $v=1/\sqrt{lc}$ and wave impedance $Z_0 = \sqrt{l/c}$, where $l$ and $c$ are inductance and capacitance per unit length respectively. In this lumped-element circuit treatment, states of the CWR and the flux qubits can be completely encompassed in terms of the flux function $\phi (x,t)=\int_{-\infty}^t V(x,t')dt'$, where $V(x,t)$ is an electrical potential of the CWR at position $x$ with respect to the surrounding ground line. The Lagrangian of the overall setup shown in Fig. \ref{CPWR}(a) is then given by
\begin{equation}
	\mathcal{L}=\sum_{j=1}^{N+1} \mathcal{L}_j ^{\text{CWR}} + \sum_{j=1}^N \mathcal{L}_j ^{\text{flux qubit}},\label{totalL}
\end{equation}  
with
\begin{equation}
	\mathcal{L}_j ^{\text{CWR}}= \sum_{k=0}^n \frac{c\Delta x}{2} (\dot{\phi}_{k}^j)^2-\sum_{k=0}^{n -1}\frac{(\phi_k^j -\phi_{k+1}^j)^2}{2l\Delta x}, \hspace{0.1cm} \text{and}
\end{equation}
\begin{equation}
	\mathcal{L}_j ^{\text{flux qubit}}=\sum_{\ell=1}^6 \frac{C_{J_\ell}}{2}\left(\dot{\widetilde{\phi}_\ell ^j }\right)^2 + E_{J_\ell}\cos \left(\frac{\widetilde{\phi}_\ell ^j}{\varphi_0} \right).
\end{equation}
Here, $C_{J_\ell}$ and $E_{J_\ell}$ are Josephson capacitance and energy of Josephson junction (JJ)$_\ell$ within the $j$th flux qubit (see Fig. \ref{CPWR}(c)). We assume that $E_{J_1}=E_{J_2}=E_{J}$, $E_{J_3}=\alpha E_{J}$, $E_{J_4}=E_{J_5}=\beta E_{J}$ and $E_{J_6}=\gamma E_{J}$, where $\alpha,\beta,\gamma < 1$. In addition, $\varphi_0 =\hbar/2e$ is the reduced flux quantum, $\widetilde{\phi}_\ell$ is flux difference across JJ$_\ell$, for example, $\widetilde{\phi}_6 =\phi_0 ^{j+1}-\phi_n ^j$ and $\Delta x$ is the lattice spacing of the lumped circuit element description. With these system parameters and the flux quantization rule, we arrive at 
\begin{eqnarray}
	\mathcal{L}_j ^{\text{flux qubit}}=&& \frac{C_J}{2}\left[\left(\dot{\widetilde{\phi}_1 ^j }\right)^2 +\left(\dot{\widetilde{\phi}_2 ^j }\right)^2 \right]+\frac{\alpha C_J}{2}\left(\dot{\widetilde{\phi}_2 ^j } -\dot{\widetilde{\phi}_1 ^j }\right)^2 +\nonumber\\ 
	&&\beta C_J \left(\dot{\widetilde{\phi}_x ^j }+\dot{\widetilde{\phi}_2 ^j } -\dot{\widetilde{\phi}_1 ^j }\right)^2 +\frac{\gamma C_J}{2}\left(\dot{\widetilde{\phi}_x ^j }\right)^2 -\nonumber\\ 
	&& {U_q}^j (\varphi_1 ^j ,\varphi_2 ^j , \varphi_x ^j ),\label{qubit_Lag}
\end{eqnarray}
where the phase slip $\widetilde{\phi}_x ^j =\phi_n ^j -\phi_0 ^{j+1}$, see Fig.~\ref{CPWR}(c), and 
\begin{eqnarray}\label{qpot}
	\frac{U_q ^j}{E_{J}}=&-&[\cos\varphi_1 ^j + \cos\varphi_2 ^j + \alpha \cos(\varphi_2 ^j - \varphi_1 ^j+ 2\pi f_1 ^j)\\
	&+&2\beta \cos (\pi f_3 ^j)\cos(\varphi_2 ^j -\varphi_1 ^j+2\pi (f_1 ^j-f_2 ^j +f_3 ^j/2)+\varphi_x ^j)].\nonumber
\end{eqnarray}
where $\varphi_k =\widetilde{\phi}_k /\varphi_0$ is a phase drop across JJ$_k$, $\varphi_0 =\hbar/2e$ is the reduced flux quantum, and $f_k =\Phi_k /(2\pi \varphi_0)$ is a frustration parameter. 

When we consider the Kirchhoff's law at the node $\phi_n ^j$, the equation of motion is given by
\begin{widetext}
\begin{eqnarray}
	ca \ddot{\phi}_n ^j +(\gamma +2\beta)C_J \left(\ddot{\phi}_n ^j -\ddot{\phi}_0 ^{j+1} \right)&&+2\beta C_J \left(\ddot{\widetilde{\phi}_2 ^j }-\ddot{\widetilde{\phi}_1 ^j } \right)=\frac{1}{l\Delta x}\left(\phi_{n-1}^j -\phi_n ^j \right)-\gamma I_c \sin \varphi_x ^j  \nonumber \\
	&&-\beta I_c  \left[\sin(\varphi_x ^j +\varphi_2 ^j -\varphi_1 ^j +2\pi(f_1 ^j -f_2 ^j))+\sin(\varphi_x ^j +\varphi_2 ^j -\varphi_1 ^j +2\pi(f_1 ^j -f_2 ^j +f_3 ^j)) \right],
\end{eqnarray}
\end{widetext}
with $I_c =E_J /\varphi_0$. From here onwards, we assume that the Josephson inductance of JJ$_6$ in each flux qubit $F^j$ is much smaller than the total inductance of each qubit loop, so that most of the current flows through the resonator. As a result, the qubit acts as a small perturbation to the CWR. With this assumption, we arrive at a simplified equation of motion
\begin{equation}
	ca \ddot{\phi}_n ^j +(\gamma +2\beta)C_J \left(\ddot{\phi}_n ^j -\ddot{\phi}_0 ^{j+1} \right)=\frac{1}{l\Delta x}\left(\phi_{n-1}^j -\phi_n ^j \right)-\gamma I_c \sin \varphi_x ^j,
\end{equation}
which is nothing but the conservation of currents at the node $\phi_n ^j$. This scenario has been thoroughly analyzed in Refs. \cite{Leib2012,Leib2014}. We thus decompose the Lagragian of JJ$_6$ into linear ($\mathcal{L}_j ^{\text{JJ lin}}$) and non-linear ($\mathcal{L}_j ^{\text{JJ nonlin}}$) components as 
\begin{equation}
	\mathcal{L}_j ^{\text{JJ lin}}=(\gamma +2\beta) \frac{ C_J}{2} \left(\dot{\phi}_{n}^j-\dot{\phi}_{0}^{j+1}\right)^2-\frac{1}{2L_J}\left(\phi_{n}^j-\phi_{0}^{j+1}\right)^2, \hspace{0.1cm} \text{and}
\end{equation}
\begin{equation}
	\mathcal{L}_j ^{\text{JJ nonlin}}= \gamma E_J \cos \left(\frac{\phi_{n}^j-\phi_{0}^{j+1}}{\varphi_0}\right) +\frac{1}{2L_J}\left(\phi_{n}^j-\phi_{0}^{j+1}\right)^2.
\end{equation}
In the continuum limit $\Delta x \to 0$, we arrive at 
\begin{equation}
	\mathcal{L}_j ^{\text{CWR}}= \int_{(j-1)a}^{ja} \left\lbrace \frac{ca}{2}[\partial_t \phi(x,t)]^2 -\frac{1}{2la}[\partial_x \phi(x,t)]^2 \right\rbrace dx, 	
\end{equation}
where $a=L/(N+1)$ is the lattice spacing between junctions JJ$_6$, that also corresponds to the whole resonator length.  
\begin{equation}
	\mathcal{L}_j^{\text{JJ lin}}=(\gamma +2\beta) \frac{ C_J}{2} \delta \dot{\phi}_j ^2-\frac{1}{2L_J}\delta \phi_j ^2, \hspace{0.1cm} \text{and}
\end{equation}
\begin{equation}
	\mathcal{L}_j^{\text{JJ nonlin}}=\gamma E_J \cos \delta \varphi_j +\frac{1}{2L_J} \delta \phi_j ^2,\label{JJnonlin}
\end{equation}
where $\delta \phi_j = \phi|_{x\rightarrow ja^{-}}-\phi|_{x\rightarrow ja^{+}}=\tilde{\phi}^j_x$ is the flux drop introduced by JJ$_6$ of the $j$th flux qubit, in the limits of the flux approaching the JJ$_6$ from its left side ($\phi|_{x\rightarrow ja^{-}}$) and from the right ($\phi|_{x\rightarrow ja^{+}}$). By considering the boundary conditions of vanishing currents at the two CWR ends: $\partial_x \phi|_{x=0}=\partial_x \phi|_{x=L}=0$, the conservation of currents at each JJ$_6$: $\partial_x \phi|_{x\rightarrow ja^{-}}=\partial_x \phi|_{x\rightarrow ja^{+}}$, and the JJ$_6$ current-flux relationship: $-\partial_x \phi|_{x=ja}/l =(\gamma +2\beta)C_J \delta \ddot{\phi}_j + \delta \phi_j /L_J$, we arrive at a well-defined eigenvalue problem \cite{Leib2014}. With solutions of the eigenmode functions, we can transform the linear part of the JJ$_6$'s doped CWR into independent harmonic oscillators \cite{Leib2012,Leib2014}. After performing a Legendre transfrom, we arrive at the full Hamiltonian 
\begin{equation}
	\mathcal{H}_{{CWR}}=\sum_i ^\infty \frac{1}{2m_i}\pi_i ^2 +\frac{1}{2}m_i ^2 \omega_i ^2 \tau_i ^2 +\mathcal{H}_{NL},
\end{equation}
where $m_i =c\int_0 ^L r_i ^2 dx +(\gamma +2\beta)C_J \sum_{j=1}^N (r_i |_{x\rightarrow ja^{-}}-r_i |_{x\rightarrow ja^{+}})^2$ is the effective mass of the $i$th eigenmode \cite{Leib2012}, $\pi_i =m_i \dot{\tau}_i$ is the canonical conjugate momentum of $\tau_i$ and $\mathcal{H}_{NL}= \frac{-\varphi_0 ^2}{L_J}\sum_{j=1}^N \left[\gamma \cos\left(\frac{\delta \phi_j}{\varphi_0} \right)+\frac{\delta \phi_j ^2}{2\varphi_0 ^2} \right]$ is the non-linear part of the Hamiltonian (c.f. Eq.  (\ref{JJnonlin})). Here, we have assumed an ansatz for the flux function $\phi(x,t)=\sum_i \tau_i (t) r_i (x)$. By imposing canonical commutation relations $[\pi_n ,\tau_m]=-i\delta_{nm}$, we quantize the theory with annihilation (creation) operators $a_i =\sqrt{m_i \omega_i /(2\hbar)}(\tau_i +i\pi_i /(m_i \omega_i))$ ($a_i ^\dagger$). Therefore, we finally arrive at $\mathcal{H}_{CWR}=\sum_i \hbar \omega_i a_i ^\dagger a_i + \mathcal{H}_{NL}$, where $\omega_i =(\pi v/L)(N+1)m$ with $m\in \mathbb{N}$, and $\nu=1/\sqrt{lc}$. We further impose that each JJ$_6$ operates in a linear approximation of Josephson inductance \cite{Bourassa2009} such that $\mathcal{H}_{NL}\approx 0$.

In a single-band approximation or a plasma frequency $\omega_p =\bar{\omega}$ with $\bar{\omega}=\pi v (N+1)/L$, when a set of $\mathcal{M}$ eigenmodes become degenerate \cite{Leib2014}, we have $\mathcal{H}_{CWR}=\sum_i \hbar \omega_i a_i ^\dagger a_i.$ In order to obtain the $j$th flux qubit energy and the qubit-resonator coupling, we expand the qubit potential term, Eq. (\ref{qpot}), up to the first order in $\varphi_x ^j$ \cite{USC_gate}. This leads us to the approximated potential energy
\begin{eqnarray}\label{QFPoT}
	\frac{U_q ^j}{E_{J}}\approx&-&[\cos\varphi_1 ^j + \cos\varphi_2 ^j + \alpha \cos(\varphi_2 ^j - \varphi_1 ^j+ 2\pi f_1 ^j)\\
	&+&2\beta \cos (\pi f_3 ^j)\cos(\varphi_2 ^j -\varphi_1 ^j+2\pi (f_1 ^j-f_2 ^j +f_3 ^j/2))]\nonumber\\
	&+&2\beta \cos (\pi f_3 ^j)\sin(\varphi_2 ^j -\varphi_1 ^j+2\pi (f_1 ^j-f_2 ^j +f_3 ^j/2))\varphi^j_x \nonumber.
\end{eqnarray}
First and second terms define the flux qubit potential, while the third term stands for the qubit-resonator coupling. The next step is to consider the numerical diagonalization of the flux qubit Hamiltonian. The latter is obtained by including the kinetic energy terms appearing in Eq.~(\ref{qubit_Lag}) and the first and second terms of Eq.~(\ref{QFPoT}), and performing a Legendre transformation. The numerical diagonalization allows us to obtain the two lowest energy levels, defining the qubit. The qubit-resonator coupling is obtained by projecting the operator $\sin(\varphi_2 ^j -\varphi_1 ^j+2\pi (f_1 ^j-f_2 ^j +f_3 ^j/2))$ into the qubit basis, that is
\begin{equation}
\sin(\varphi_2 ^j -\varphi_1 ^j+2\pi (f_1 ^j-f_2 ^j +f_3 ^j/2))=\sum_{\nu=0,x,y,z}c_\nu\sigma_\nu, 
\end{equation}
with $\sigma_0=\mathbbm{1}$ being the identity operator, and $c_\nu$ are $c$-numbers obtained numerically. Hence, the Hamiltonian of the overall setup $\mathcal{H}=\mathcal{H}_{CWR}+\mathcal{H}_{\text{flux qubits}}+\mathcal{H}_{\text{interaction}}$ (c.f. Eq. (\ref{totalL})) becomes
\begin{equation}
\mathcal{H} = \frac{\hbar}{2}\sum^{N}_{j=1}\omega^j_q\sigma^j_z+\hbar\sum_{\ell\in\mathcal{M}}\omega_{\ell}a^{\dag}_{\ell}a_{\ell}+ \hbar\sum^{N}_{j=1}\sum_{\ell\in\mathcal{M}}g_j(c^j_x\sigma^j_x+c^j_z\sigma^j_z)(a_{\ell}+a^{\dag}_{\ell}),\nonumber
\end{equation}
which is the starting point of the main text, Eq. (\ref{effH}). Here, $g_j=2\beta E_J \cos(f_3)\delta \phi_j$ are the effective coupling strengths between the flux qubits and the CWR at the degeneracy point with $\delta \phi_j \propto\sqrt{2/(N+1)}\sin (p_j)$ with $p_j =\pi j/(N+1)$ \cite{Leib2014}.

\section{Derivation of the evolution operator}
It has been shown that magnetic fluxes $\phi_1 ^j$ can tune the coefficients $c_x ^j$ and $c_z ^j$ (see Fig.~2(a,b) of the main text and Ref. \cite{USC_gate}) to arrive at the longitudinal coupling with $c_x ^j \approx0$ and $c_z ^j\approx1$ \cite{USC_gate,Peropadre2010}, which is an ideal condition for the pairwise cluster state generation in an ultrafast timescale. For each mode $\ell$, we define a displacement operator
\begin{equation}
	\mathcal{D}_{\ell} (\sum_j \kappa_j \sigma_z ^j)=\exp [(\sum_j \kappa_j \sigma_z ^j )a_{\ell} ^\dagger -(\sum_j \kappa_j \sigma_z ^j )a_{\ell} ],
\end{equation}
with $\kappa_j =g_j /\omega$ and $\omega_{\ell} =\omega$ since we consider a collective resonator mode at a degeneracy point \cite{Leib2014}. In addition, for all the modes within the manifold $\mathcal{M}$, we define a collective displacement operator 
$
	\mathcal{D}(\xi)=\prod_{{\ell}\in M} e^{\xi a_{\ell} ^\dagger -\xi^\ast a_{\ell} },
$
where $\xi=\left(\sum_j \kappa_j \sigma_z ^j \right)$. By transforming the original Hamiltonian, Eq.  (\ref{effH}) with the above operator, we obtain 
\begin{equation}
H=\mathcal{D}^\dagger (\xi)\mathcal{D}(\xi)H\mathcal{D}^\dagger (\xi)\mathcal{D}(\xi)=\mathcal{D}^\dagger (\xi) [\omega \sum_{\ell} a_{\ell} ^\dagger a_{\ell} -\omega M\xi^2 ]\mathcal{D} (\xi),
\end{equation}
where $M$ is the dimension of $\mathcal{M}$. The associated evolution operator is given by 
\begin{eqnarray}
	U(t)&=&U_0 (t) e^{i\omega t M \xi^2}e^{-i\omega t \mathcal{D} ^\dagger (\xi)(\sum_{\ell} a_{\ell} ^\dagger a_{\ell})\mathcal{D} (\xi)}\\
	&=&U_0 (t) e^{i\xi^2 M(\omega t - \sin(\omega t))}\prod_{\ell} e^{-i\omega t a_{\ell} ^\dagger a_{\ell}} \mathcal{D}_{\ell}\big[\xi(t)\big],\nonumber
\end{eqnarray}
with $U_0 (t)=\exp[{-it \sum_j \frac{\omega_q ^j}{2}\sigma_z ^j}]$ and $\mathcal{D}_{\ell}\big[\xi(t)\big]=\mathcal{D}_{\ell}((1-e^{i\omega t})\xi)$. After an evolution time $t=2\pi n/\omega$,
\begin{equation}
U(2\pi n/\omega)=U_0 (2\pi n/\omega) e^{i\xi^2 M(2\pi n)}\prod_{\ell} e^{-2\pi  n i a_{\ell} ^\dagger a_{\ell}},
\end{equation}
where $n$ is an integer multiple. Since our protocol constitutes of pairwise qubits, 
\begin{eqnarray}
U(2\pi/\omega)\approx &&  \exp [\frac{-i\pi}{\omega}(\omega_q ^i \sigma_z ^i +\omega_q ^j \sigma_z ^j)\times\nonumber\\
	&& \exp [i 4\pi nM ((\kappa_i ^2 +\kappa_j ^2)\frac{\mathbbm{1}}{2}+\kappa_i \kappa_j \sigma_z ^i \sigma_z ^j ) ].
\end{eqnarray}
Thus, we have
\begin{eqnarray}
U_{\mathcal{CZ}}=&& \mathcal{U}\times \exp\left[\frac{-i\pi}{4}(\sigma_z ^i +\sigma_z ^j) \right]
\times\nonumber\\
&& \exp\left[4\pi i M\left((\kappa_i ^2 +\kappa_j ^2)\frac{\mathbbm{1}}{2}+\kappa_i \kappa_j \sigma_z ^i \sigma_z ^j \right) \right],
\end{eqnarray}
where $\mathcal{U}=\exp \left[\frac{-i\pi}{4} \left[\Big(\frac{4\omega_q ^i -\omega}{\omega}\Big)\sigma_z ^i +\Big(\frac{4\omega_q ^j -\omega}{\omega}\Big)\sigma_z ^j \right]\right]$ and $n=1$. To perform the controlled phase gate operation with a maximum fidelity, we require that both $\kappa_i ^2 +\kappa_j ^2 =\frac{1}{8nM}$ and $\kappa_i \kappa_j =\frac{1}{16nM}$ are satisfed. In other words, we need $\kappa_i + \kappa_j =\frac{1}{2\sqrt{nM}}$.



\begin{thebibliography}{9}
\bibitem{supplemental} See Appendix, which provides a detailed analysis of the lumped circuit element of Fig.~\ref{Fig1}(c), the derivation of Hamiltonian, Eq. (\ref{effH}), and the evolution operator $U_{\mathcal{CZ}}(t)$.
\end{thebibliography}
\end{document}